\documentclass{elsart}
\usepackage{natbib}

\usepackage{graphicx}

\usepackage{amssymb}

\begin{document}
\newcommand{\chandra}{{\it Chandra}}
\newcommand{\xmm}{$XMM$-$Newton$}
\newcommand{\rosat}{{\it ROSAT}}
\newcommand{\einstein}{{\it Einstein}}
\newcommand{\brems}{bremsstrahlung}
\newcommand{\tetp}{$T_{\rm e}/T_{\rm p}$}
\newcommand{\teti}{$T_{\rm e}/T_{\rm ion}$}
\newcommand{\arcsecond}{$^{\prime\prime}$}
\newcommand{\ion}[2]{{#1} {\textsc{#2}}}

\begin{frontmatter}



\title{Electron-Ion Temperature Equilibration at Collisionless Shocks
  in Supernova Remnants}


\author{Cara E. Rakowski}
\ead{crakowski@cfa.harvard.edu}
\address{Harvard-Smithsonian Center for Astrophysics,
60~Garden~St. MS-70, Cambridge, MA 02138, USA}

\begin{abstract}

The topic of this review is the current state of our knowledge about
the degree of initial equilibration between electrons, protons and
ions at supernova remnant (SNR) shocks. Specifically, the question has
been raised as to whether there is an inverse relationship between the
shock velocity and the equilibration similar to the relationship
between equilibration and Alfv\'en Mach number seen in interplanetary
shocks \citep{Schwartz88}. This review aims to compile every
method that has been used to measure the equilibration and every SNR
on which they have been tested.  
I review each method, 
its problems and uncertainties and how those would effect the degree
of equilibration (or velocity) inferred. The final compilation of
observed electron to proton temperature ratios as a function of shock
velocity gives an accurate, conservative picture of the state of our
knowledge and the avenues we need to pursue to make progress in our
understanding of the relation between the velocity of a shock and the
degree of equilibration. 

\end{abstract}


\end{frontmatter}

\section{Introduction}

The nature of electron and ion heating behind collisionless shocks
remains an open question in shock physics.  While the overall kinetics
of a shock can be described simply in terms of the Rankine-Hugoniot
solutions to the equations for conservation and continuity \citep[see
e.g.][]{McKee80}, the fraction of the shock's kinetic energy that is
transferred to the thermal and cosmic-ray populations of electrons and
ions is unknown. 
Theoretically, anything 
between $(T_{e}/T_{p})_{0} \sim m_{e}/m_{p}$ to rapid full 
equilibration has been proposed. However mounting observational
evidence points to an intermediate degree of initial electron  
heating, where the disparity between electron and ion heating
increases with shock velocity. This review examines these
measurements in order to point the way towards further progress.

Modeling non-linear acceleration of particles in high Mach number
collisionless shocks remains a challenge. There are numerous
analytically derived instabilities that could be important, but it is
not necessarily straight-forward to interpret particle simulations 
in terms of these processes. While comparing the simulations to observations
of solar wind shocks has been quite successful, the results may not
have a strong bearing on the important processes in supernova remnant
(SNR) shocks which are significantly faster and less dense. For a recent,
thorough review of the open problems in collisionless-shock physics
see \citet{Lem04}. For the purpose of this review, it is important to
keep in mind that the electro-magnetic waves that define these
instabilities and the shock itself will affect the ions and electrons
differently due to the large inertial difference. 

Given the current ambiguity regarding the partitioning of energy at a
shock, interpretations of SNR observations commonly assume purely
thermal particle distributions and consider the extreme cases of
minimal equilibration (mass-proportional heating) or full
equilibration (all species rapidly heated to the same temperature). 
\footnote{\citet{Bocchino99} turned this argument around to estimate a
  degree of equilibration based on given Sedov parameters for the Vela
  supernova remnant}   
If cosmic ray production is included, the case considered is usually
one with an extremely high injection efficiency of a thermal seed
population into the cosmic ray acceleration process so that the
cosmic ray flux at Earth can be explained.
For observations that only measure one or two populations, the choice of
equilibration can drastically affect the inferred age, explosion energy
or distance.
Furthermore, if equilibration is a function of shock speed,
then even spanning both scenarios may be misleading because the
evolution of the initial shock equilibration will effect the dynamics
of the remnant (for instance consider the internal pressure). 
Calibrating a relationship between shock velocity and the efficiency
of heating different particles to thermal and cosmic ray energies
would be a fantastic boon for SNR studies if a clean relationship
exists.

There are three generic methods for determining the temperature of
particles behind a shock: (1) \brems\ continuum emission, (2) the
thermal broadening of a line from a particular species, and (3) the flux
ratios of lines of either a single element or between elements. Each
of these methods has its own pitfalls.
(1) Both thermal bremsstrahlung and synchrotron radiation (from a high
energy population of electrons) produce continuum emission in the
X-rays, and their relative contributions are not always possible to
disentangle. Furthermore the \brems\ continuum in the X-rays samples 
emission from an electron
population significantly downstream from the shock. Assuming the
electrons and protons were out of equilibrium the temperatures
will equilibrate downstream on the timescale of coulomb
collisions. The initial electron temperature must then be inferred
using the ionization state to calibrate the coulomb timescale post-shock.
(2) The line broadening is a function not only of the temperature of that
species but also the velocity structure of the shock if the shocked
filament is not resolved. 
(3) The temperature dependence of ionization and excitation by
electrons and protons can be used to diagnose $T_{\rm e}$ or 
$T_{\rm p}$ from the line flux ratios. The fluxes also depend on the
ion fractions or relative abundances which must be assumed if they
cannot be derived.
We will examine many manifestations of these generic methods and
problems as the observational evidence unfolds. 

The body of this review is organized by supernova in order of
increasing velocity to emphasize the necessity of measurements in
multiple wave-bands to corroborate the degree of
equilibration. Although different authors use different parameters to
express the degree of equilibration, the ratio of the initial electron
to proton temperatures at the shock front (\tetp )  will be used
throughout this review.  
These studies all concentrated on the equilibration at the outer blast
wave of each remnant. Further complications must be considered with
regard to the energetics of a reverse shock into the pure metal ejecta,
where the energetics are not dominated by the protons, cold electrons
are continually released by ionization, and turbulence may generate a
high magnetic field. 

\vspace*{-12pt}
\section{The Cygnus Loop and RCW 86 \label{s_cyg}}

\vspace*{-12pt}
The ratio of broad-to-narrow component flux in the H$\alpha$ and
H$\beta$ lines was used by \citet{Ghav01} to measure the
electron-proton temperature ratio behind shocks in the Cygnus Loop and
RCW 86. The theory behind the broad and narrow Balmer line emission
was first laid out by \citet{CRK80}.
Directly behind a collisionless shock there can exist a population of
cold neutral ions that were not affected by the shock passage. These
neutrals will emit H$\alpha$ and H$\beta$ Balmer lines if they are
collisionally excited before being ionized. This produces a narrow
component of emission, whose width reflects the temperature of the
pre-shock gas. The cold neutrals can also charge exchange with the
shock-heated protons resulting in a population of hot neutrals. A
broad component to the Balmer lines arises from this hot neutral
population, where the width of the broad line represents the
temperature of the post-shock protons.

\begin{figure}
\includegraphics[scale=0.6, keepaspectratio]
{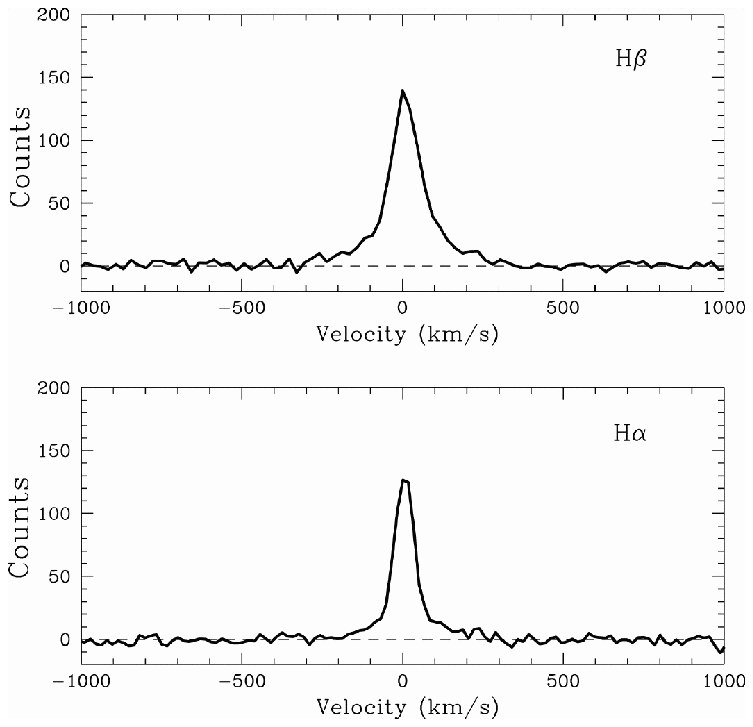}\hfill
\includegraphics[scale=0.6, keepaspectratio]
{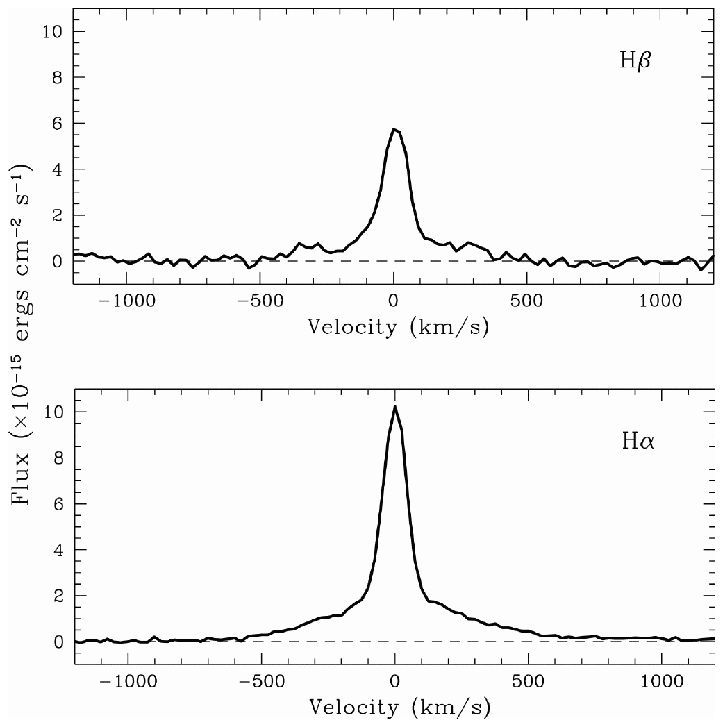}\hfill
\includegraphics[scale=0.6, keepaspectratio]
{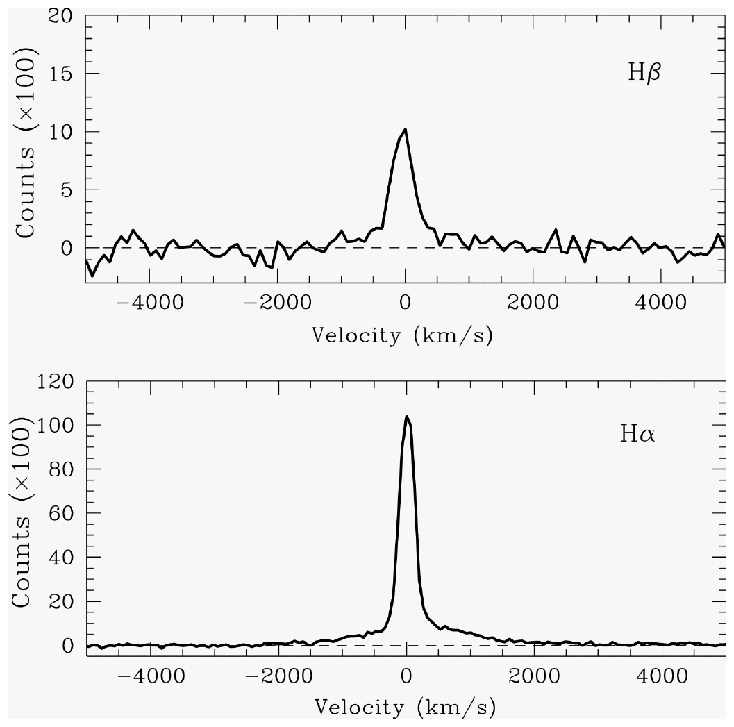}
\caption{The H$\alpha$ and H$\beta$ profiles of (a) the Cygnus Loop,
(b) RCW 86 and (c) a compact knot in Tycho's SNR from
\citet{Ghav01}.
Note that the H$\alpha$ profile has higher signal to
noise to constrain the broad-to-narrow ratio than does H$\beta$, with
the possible exception of the Cygnus Loop.}
\label{ha_spec}
\end{figure}

\begin{figure}
\includegraphics[scale=0.6, keepaspectratio]
{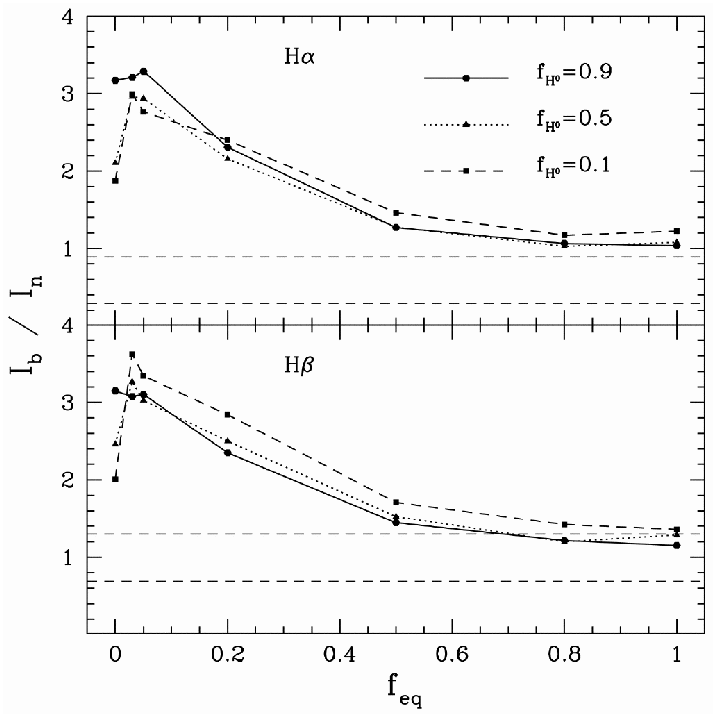}\hfill
\includegraphics[scale=0.6, keepaspectratio]
{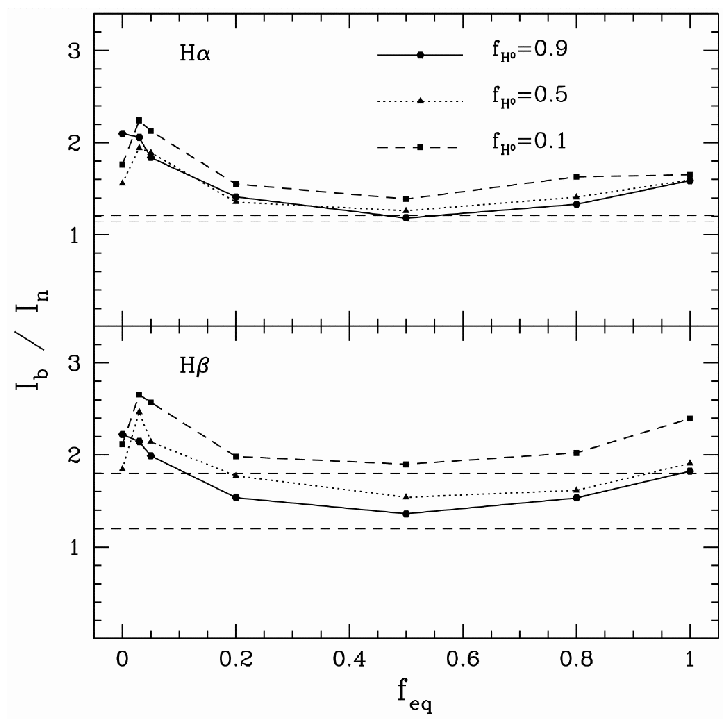}\hfill
\includegraphics[scale=0.6, keepaspectratio]
{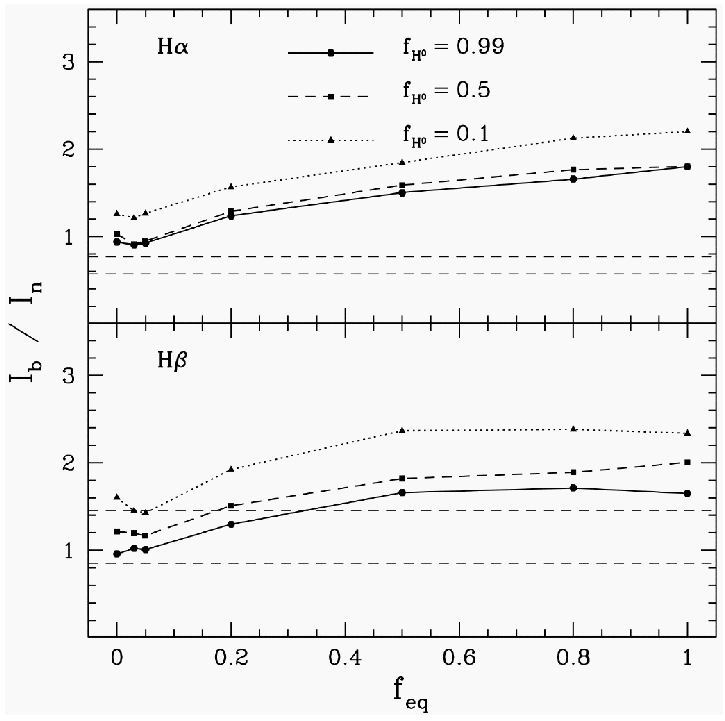}
\caption{The predicted H$\alpha$ and H$\beta$ broad-to-narrow ratios
as a function of ``fractional equilibration'' from \citet{Ghav01} for
velocities appropriate for (a) the Cygnus Loop, (b) RCW 86 and (c)
Tycho.
Note that the model predictions are inconsistent with the higher
signal-to-noise H$\alpha$ broad-to-narrow ratio for the Cygnus Loop
and Tycho's SNR.}
\label{IbIn}
\end{figure}

Figures \ref{ha_spec}a and \ref{ha_spec}b from \citet{Ghav01} 
present the H$\alpha$ and H$\beta$ profiles from the Cygnus Loop and
RCW~86.\footnote{The data are presented without the model so that the 
reader can better evaluate the reliability of the broad component
measurement for each remnant in both H$\alpha$ and H$\beta$.}
The post-shock proton temperatures derived from the widths of
the H$\alpha$ broad components are $T_{p}=(1.5\pm0.26) \times
10^{6}$ K for Cygnus and $T_{p}=(6.9\pm0.2) \times 10^{6}$ K for
RCW~86. To estimate the degree of equilibration between electrons and 
protons \citet{Ghav01} modeled the broad and narrow emission
including the temperature dependence of charge exchange and excitation
of the fast  and slow neutrals by electrons and protons. 
The predicted broad-to-narrow ratios are plotted for
different values of the initial neutral fraction in 
figures \ref{IbIn}a and \ref{IbIn}b from \citet{Ghav01}.
Using the H$\beta$ ratio for Cygnus and the H$\alpha$ ratio for
RCW~86, they report constraints of \tetp = $0.67-1.0$ and \tetp =
$0.25-0.43$, respectively. 
Assuming cosmic ray production was unimportant to the
energetics of the shock, these equilibrations and the broad component
widths imply shock speeds of 300-400 km s$^{-1}$ (Cygnus) and 580-660
km s$^{-1}$ (RCW~86).

However, it is worth noting from figure \ref{IbIn}a 
that the H$\alpha$ broad-to-narrow ratios
predicted by \citet{Ghav01}  for the Cygnus loop lie 10\% higher than
those observed. This level of discrepancy is readily explained by
contamination of the narrow component by emission from a shock
precursor.  
Some fraction of the hot neutrals will escape ahead of the shock
before being ionized. 
Furthermore, in a cosmic-ray modified shock, some fraction of the
cosmic rays will also escape into the pre-shock gas.
The hot neutrals and cosmic ray particles
will ionize and excite the medium ahead of the shock. This non-linear
effect was not included in the models of \citet{Ghav01} and in fact,
modeling the balance between the production of hot neutrals and the
continued existence of neutral atoms at the shock front is a work in
progress (K. Korreck, personal communication). 
However, it is clear that precursor emission will
increase the narrow component flux, shifting all the models to lower
broad-to-narrow ratios. Given this uncertainty it is unclear
whether a measured broad-to-narrow ratio lower than all of the models 
provides any constraint on \tetp\ at all. In fact in sections
\ref{s_deml71} and \ref{s_tycho} we will see examples of shocks in
which the broad-to-narrow ratio is significantly lower than any of
Ghavamian's models, strongly indicating precursor emission. At present
there is no reason to assume that this level of precursor
contamination is missing in the Cygnus loop or RCW~86 just because
their broad-to-narrow ratios happen to marginally intersect the
predicted curves. 

On the other hand, there is corroborating evidence for a high degree
of equilibration in the Cygnus Loop. \citet{Ray03} examined the far
ultraviolet spectrum at the same position as was studied in
\citet{Ghav03}. The width of the \ion{O}{iv} $\lambda$1032 line gives an
upper limit on the oxygen temperature of $2.7 \times 10^{6}$ K (any velocity
structure broadening would decrease this estimate). This is within a
factor of 2.5 of the hydrogen temperature measured in \citet{Ghav01}.  
Given that mass-proportional heating would lead to a factor of 16
difference between the temperatures of oxygen and hydrogen, this
measurement indicates a high degree of equilibration between ions at
the shock. This lends credibility to the claim of high electron-ion
equilibration. However, while the electron heating mechanisms may be
tied to the temperature of the ions, there is still a three orders of
magnitude difference in mass. Hence a high degree of equilibration
between the ions themselves does not necessarily imply a similarly
high degree of equilibration between the electrons and ions.

\section{SNR DEM L71 \label{s_deml71}}

A combination of optical and X-ray observations was required to
constrain the electron-proton temperature ratio around the blast-wave
of SNR DEM~L71 \citep{Ghav03,Rak03}. 
This Large Magellanic Cloud (LMC) remnant has many non-radiative
H$\alpha$ filaments around its outer rim, ranging in velocity from 
$\sim 500 - 1000$ km s$^{-1}$. \citet{Ghav03} and \citet{Rak03}  chose
apertures along the blast-wave to combine areas of similar optical
properties into large enough regions for meaningful spectra in a
45.4 ks \chandra\ observation. 
The optical measurements of the H$\alpha$ broad component width (to
obtain $T_{\rm p}$) and attempts to model the broad-to-narrow ratio
are presented in \citet{Ghav03}. 
Without the addition of precursor
emission the predicted broad-to-narrow ratios never drop below 1.0,
but the measured flux ratios are never more than 0.74 (generally
around 0.5). Until full precursor modeling is available no
conclusion can be made about the degree of equilibration purely from
the optical data. 

\begin{figure}
\centerline{
\includegraphics[scale=1.0, keepaspectratio]
{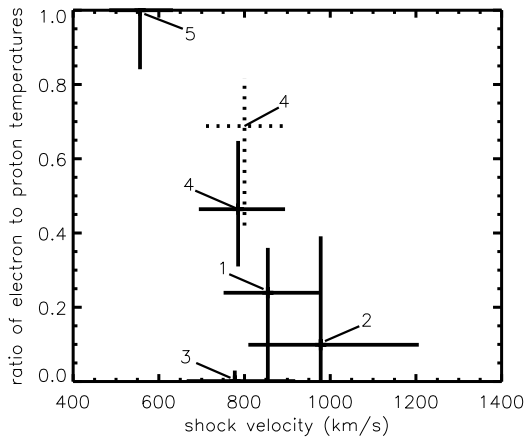}
}
\caption{Initial electron to proton temperature ratio as a function of
shock speed for five apertures in DEM L71. Published results
\citep{Rak03} are plotted as solid crosses, new result as dashed. The
new analysis incorporates 100ks more data which enables constraints on the
abundances for each region. The offset between the old and new results
illustrates how an assumption about the underlying abundance set can
bias the measured temperature.}
\label{dem_vg0}
\end{figure}

In \citet{Rak03}, we extracted \chandra\ X-ray spectra from three
nested regions behind the blast-wave at each aperture to measure the
evolution of the electron temperature.
Our final analysis of the initial ratio, \tetp , utilized a planar
shock model with the initial proton temperature fixed to the values
from \citet{Ghav03}. The electron temperature as a function of
distance behind the shock could then be parameterized in terms of the
initial degree of equilibration.
The published
results are shown as solid crosses in figure \ref{dem_vg0}. 
Except for the slowest shock (region 5), all of the shocks were
consistent with no equilibration in the preliminary tests. Combining
X-ray and optical results allowed us to identify one shock (region 4)
with \tetp $\sim 1/2$ whose spectrum was inconsistent with either full
or minimal equilibration given the measured proton temperature.
Limited statistics dictated that the
abundances and column density be held constant in the X-ray fits. 
The chosen average abundance set will introduce a bias  given the
inverse correlation between the abundance and temperature when both are
allowed to vary. 
The errors shown include an estimate of this
systematic error as well as the measurement uncertainty in $T_{\rm p}$.
An additional 100 ks of \chandra\ observing time has allowed us to
eliminate this systematic uncertainty by measuring the
abundances and column density at each aperture. 
These data confirm that our shock model is sound, i.e. the
temperatures and ionization timescales of the nested regions do indeed
evolve when fit separately.
The first of the updated measurements of \tetp\ is shown by the dashed
cross in figure \ref{dem_vg0} confirming an intermediate degree of
equilibration for region 4. 
While the uncertainty remains similar, the result
itself is no longer biased by an arbitrary choice of abundances.

\section{Tycho's SNR \label{s_tycho}}

H$\alpha$ and H$\beta$ emission from a compact knot in Tycho's SNR were
also included in \citet{Ghav01} (see section \ref{s_cyg}). 
The spectra and model predictions for the
broad-to-narrow ratios are shown here in figures \ref{ha_spec}c and
\ref{IbIn}c from \citet{Ghav01}. 
The H$\alpha$ profile has considerably better
signal to noise than H$\beta$, but the H$\alpha$ broad-to-narrow ratio
falls below predictions even after accounting for some diffuse
emission above the knot.  
\citet{Ghav01} therefore use the less well-constrained H$\beta$ ratio to
estimate \tetp\ $<$ 0.1. 
The predicted broad-to-narrow ratios in figure \ref{IbIn}c
certainly favor a low equilibration. The narrow component
contribution from the cosmic-ray precursor would have to be more than
one third of the total narrow H$\alpha$ emission in order to be
consistent with a \tetp\ ratio greater than 0.2.  
However, a more precise statement on \tetp\ awaits a full precursor
model to understand the difference between the predicted and observed
H$\alpha$ ratio.

\citet{Hwang02} studied the Tycho SNR blast-wave with \chandra . 
Spectra extracted from narrow regions along the outer rim exhibited
little or no line emission. Assuming this emission is thermal in origin 
a very young ionization age is needed to explain the lack of line
emission given a fitted electron temperature around 2 keV. 
Given the low timescale, the
electron temperature they measure should not have evolved since the
initial shock heating. This electron temperature can be compared with
a shock velocity derived (at a distance of 2.3 kpc) from radial
expansion measurements in the radio \citep[3000--4000 km
s$^{-1}$][]{Rey97} and X-rays \citep[$4600\pm400$ km s$^{-1}$][]{Hughes00a}
to derive the ratio of the electron to mean temperatures. 
There remains a debate over which expansion
measurement is representative of the present shock speed, however
the electron-to-mean temperature ratio would be 0.10$-$0.20 assuming
the mean radio expansion or 0.054$-$0.10 for the X-ray expansion
speed. Combining both these ranges yields a limit on \tetp\ of 0.03$-$0.12. 
This is in good agreement with the results of \citet{Ghav01},
however at a much faster shock within the same
remnant. \citet{Hwang02} note that the low measured electron
temperature does not require a significant cosmic ray population to
explain the gap between the mean temperature for the measured shock
speed and the electron temperature. However, neither does it exclude the
possibility that a significant fraction of the shock energy has gone
into accelerating cosmic rays.

The above conclusions were derived under the assumption that the
blast-wave emission is thermal.  However, synchrotron emission arising
from high energy electrons accelerated to cosmic-ray energies at the
shock front is an equally natural explanation for such a featureless 
spectrum. The presence of hard X-ray emission up to 30 keV requires
that a substantial portion of the 0.5-10 keV continuum be nonthermal.
(see section 4.2 of \citet{Hwang02} for a full discussion) 
In fact, \citet{Hwang02} were able to successfully model the line-free rim 
spectra with a synchrotron emission model that required consistency with radio
measurements. 
They found that if the emission is primarily non-thermal, any
additional thermal component would have to have a considerably lower
electron temperature than was found assuming a completely thermal origin.
This would be consistent with a picture where a large portion of the
shock energy has gone into accelerating cosmic rays, leaving behind a
low temperature population of thermal electrons, which may or may not
be equilibrated with the thermal population of ions.

\section{SN 1006, fully unequilibrated shocks \label{s_sn1006}}

SN 1006 is the one remnant where the equilibration between particles
has been truly well-studied. Four lines of evidence all indicate a low
degree of equilibration or nearly mass-proportional heating. Any one
estimate of \tetp\ may not be convincing, either because of its
indirectness \citep{Ray95,Vink03} or its dependence on arguments
regarding the abundances or ion fractions \citep{Ghav02,Laming96}. 
However, the convergence of four largely independent measurements is 
compelling.   

\citet{Ray95} observed the ultraviolet spectrum from a
197\arcsecond\ long region along the north-west shock measuring
emission lines from \ion{H}{i}, \ion{He}{ii},  \ion{C}{iv}, \ion{N}{v},
and \ion{O}{vi}. They found that lines from all species show similar
widths consistent with a speed of 2300 km s$^{-1}$ indicating that
ion-ion equilibration is ineffective. In \citet{Laming96}, they
modeled the line ratios for this ultraviolet spectrum. From the ratio
of two lines, one excited primarily by electrons, \ion{He}{ii}
$\lambda$1640, and one excited by protons, \ion{C}{iv} $\lambda$1550,
one can infer the ratio of electron to ion temperatures. Assuming
solar abundances and a factor of two carbon depletion onto grains,
this flux ratio indicates \teti $<0.05$. However, since the flux ratio
is obviously dependent on the assumed relative abundances, the authors
admit that a ratio of \teti $=0.2$ cannot be ruled out. 

\citet{Ghav02} re-examined the optical spectrum of the northwestern
shock. The most useful  diagnostic pair of lines were \ion{He}{i}
$\lambda$6678 and \ion{He}{ii} $\lambda$4686.  The ratio of these two
lines depends on the initial neutral fraction and the ratio of the
\ion{He}{i} excitation to ionization rates. 
The excitation to ionization rate drops sharply
above $T_{\rm e} > 10^{5}$~K, implying that a low electron temperature is
needed to produce measurable \ion{He}{i} emission. A high neutral
fraction is also necessary to replenish the supply of \ion{He}{i}
which is still ionizing faster than it emits photons.
Quantitatively, comparing the electron temperature derived from
modeling the \ion{He}{i} to \ion{He}{ii} ratio with
the proton temperature measured with the
H$\alpha$ broad component width,  yields a limit of \tetp $< 0.03$. 
The other measured line ratios, \ion{H}{i}/H$\alpha$ and the
broad-to-narrow ratios of H$\alpha$ and H$\beta$, while
less directly diagnostic, are all consistent with a low equilibration,
\tetp $< 0.07$. 

\citet{Vink03} studied the X-ray grating spectrum of a compact knot
in the northwest using \xmm . The broadening of a complex of
\ion{O}{v}, \ion{O}{vi} and \ion{O}{vii} lines imply an oxygen
temperature of $kT_{\rm oxygen} = 530 \pm 150$~keV. 
If there has been no ion equilibration, as suggested by
\citet{Ray95}, this oxygen temperature translates into a shock
velocity of $v_{s} > 4000$ km s$^{-1}$. If ion equilibration has
occurred that would require an even faster, more energetic shock.
\citet{Vink03} use a CCD-resolution X-ray spectrum, extracted from the
same \xmm\ observation, to measure an electron temperature of 
$kT_{\rm e} \sim 1.5$~keV. 
The ratio of the electron and oxygen temperatures imply an
electron-to-proton temperature ratio of $\sim$0.05 for the case of no ion
equilibration, or even smaller values for a shock where the ions are
equilibrated. 

All of the measurements to date of SN~1006 indicate that the electrons
and ions are far from being in temperature equilibrium. 
However, one of the most remarkable things about SN 1006 is the
incomplete shell of bright radio and X-ray synchrotron emission in the 
northeast and southwest indicating sites of efficient cosmic ray
acceleration. All measurements concerning equilibration
have been made for the northwestern rim, which does not
seem to emit X-ray synchrotron emission.
The situation for the synchrotron rims may be quite different.

\section{SN 1987A \label{s_1987A}}

A single \chandra\ grating spectrum from an unresolved SNR, such as SN
1987A has the potential to measure the ion-ion and electron-ion
equilibration simultaneously. The line ratios of individual elements
can be used to constrain the electron temperature, while the line
widths are a function of the temperature of that element and the
profile of its bulk motion. The \chandra\ high energy transmission
grating (HETG) spectrum of SN 1987A, presented by \citet{Michael02},
displays K$\alpha$ and L$\alpha$ lines of N, O, Ne, Mg, and
Si. Constraints on $kT_{\rm e}$ and the ionization timescale from line
ratios of individual elements were shown in \citet{Michael02}.
Unfortunately the grating spectrum is not deep enough to constrain the
temperature and timescale fully, however the line ratios agree with
CCD spectrum best-fit values of $kT_{\rm e} \sim 2.6$~keV and $n_{\rm
e}t \sim 6 \times 10^{10}$ cm$^{-3}$ s. 

\begin{figure}
\centerline{
\includegraphics[scale=1.0, keepaspectratio]
{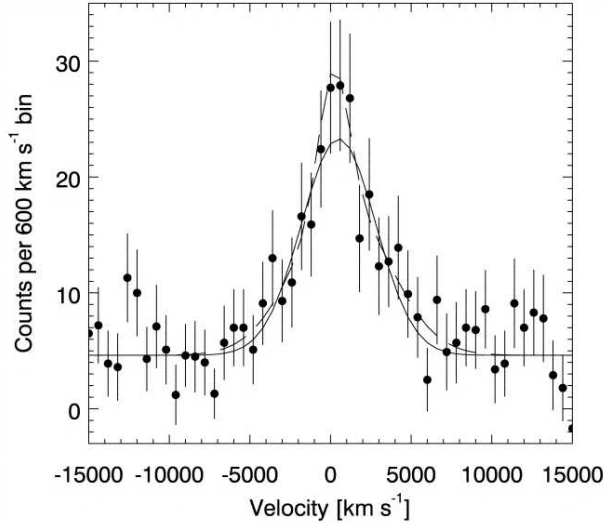}
}
\caption{Combined X-ray line profile from the grating spectrum of SN
1987A, reproduced from \citet{Michael02} figure 8. Single (solid) and
double (dashed) Gaussian fits are over-plotted.}
\label{SN87A_ionline}
\end{figure}

\begin{figure}
\centerline{\includegraphics[scale=1.0, keepaspectratio]
{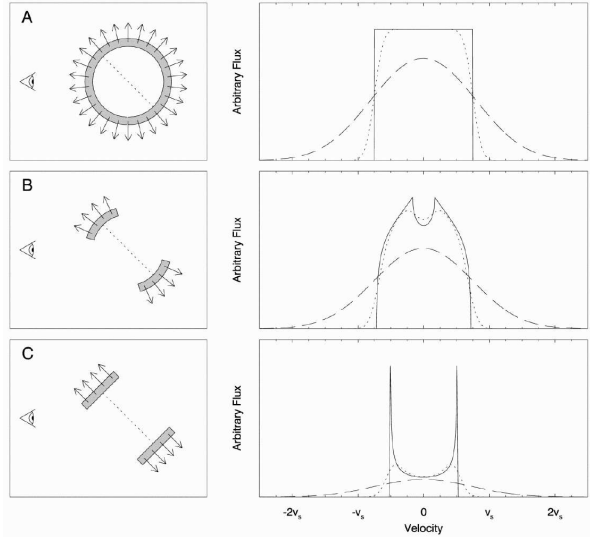}}
\caption{Line profiles predicted for different geometries and degrees
of equilibration for SN 1987A, reproduced from figure 9 of
\citet{Michael02}. Solid lines show the bulk motion profiles. The
dotted lines include the thermal broadening assuming full ion-ion
equilibration. The dashed lines assume minimal equilibration at the
shock front.} 
\label{SN87A_geom}
\end{figure}

Ideally one would like to constrain each of the line widths
separately, but again the statistics do not allow this. However, given
that results from SN 1006 indicate a low degree of ion equilibration
at such fast shocks, it is not unreasonable to assume that all the
ions have the same velocity. Therefore, \citet{Michael02} co-added the
velocity profiles of the brightest lines of N, O, Ne, Mg, and
Si to measure an ``average'' line-width. 
The combined X-ray line profile is reprinted here in figure
\ref{SN87A_ionline}. 
Since SN1987A is not resolved by the grating, this line profile
includes both thermal velocity and bulk motion.  
Figure \ref{SN87A_geom} \citep{Michael02} 
illustrates the line profiles produced by different geometries of the X-ray
emitting plasma. Full equilibration between the ions produces sharper
line profiles than no equilibration because of the greater bulk motion
contribution. The smooth, no equilibration cases are more similar to
the line profile seen. \citet{Michael02} note that equatorial
geometries are favored by other observations. 
For no ion equilibration and equatorial geometries
$v_{s} = 3400 \pm 700$ km s$^{-1}$, consistent with radio expansion
measurements. 

Similarly to what was done in DEM L71 by \citet{Rak03}, 
the velocity of the shock in SN 1987A can be used to constrain the CCD
spectrum in terms of the initial electron to ion temperature
ratio. \citet{Michael02} find that if $v_{s} \sim 3500$ km s$^{-1}$,
then the CCD spectrum is fit best by $T_{\rm e}/T_{s}= 0.11 \pm 0.02$
(where $T_{s}$ is the average shock temperature, equivalently 
\tetp $= 0.07 \pm 0.01$).  
Unlike \citet{Rak03}, these
constraints assume a fixed velocity and do not account for the range
in velocities allowed by their line width calculations. However, the
authors do note that fits with a very low electron temperature 
are inconsistent with the data, thus excluding the case of zero
election-ion equilibration. 


\section{1E 0102$-$72 \label{e0102}}

Up until this section, all discussions have assumed that the
acceleration of cosmic rays is not an important factor in the shock
dynamics. However it is widely believed that SNR shocks are the main
source of the bulk of the cosmic ray distribution. 
If cosmic ray acceleration is efficient a significant portion of the
shock energy may be diverted from the thermal populations. This should
have important consequences on the shock itself \citep{DEC00}.
The shock compression will rise due to the lower adiabatic index of
relativistic particles, the forward and reverse shocks will be closer
together and  the temperature and density gradients will be more
severe. As cosmic rays escape the shock they carry away shock
energy, making these dynamic effects more dramatic.  Neglecting the
energetics of cosmic ray acceleration when considering the
equililbration of the thermal populations can give misleading results.
If cosmic ray
acceleration is important then results that are based on an expansion
velocity and an electron temperature may in fact have a much higher
degree of equilibration between the thermal populations \citep[i.e. the
Tycho results of][]{Hwang02}. Furthermore, 
inferring a velocity purely from the electron and ion temperatures 
will underestimate the shock speed if much of the shock energy is in
relativistic particles. 

\citet{Hughes00b} used \chandra\ calibration observations of
1E~0102$-$72 to measure the expansion of the SNR since the earlier
\einstein\ and \rosat\ images. With the known 50~kpc distance to the
LMC, the 0.1\% expansion per year implies a $\sim$6000 km s$^{-1}$
shock speed for the outer blast-wave. An X-ray spectrum of the
blast-wave, extracted from the same observation, exhibits an electron
temperature of less than 1~keV, 25 times smaller than the minimum
average temperature implied by the measured shock velocity. There is
no initial \tetp\ ratio for which the gulf between this average
temperature and the electron temperature can be explained. Even after
accounting for adiabatic decompression and only coulomb heating
post-shock (up to the ionization timescale seen in the spectrum) the
minimum electron temperature for such a fast shock is no lower than
$\sim$2.5~keV for the extreme case of mass-proportional heating. 
The discrepancy between the low
electron temperature and high energy of the shock requires the
existence of another population into which the shock energy is
flowing. If a significant fraction of the shock energy has gone into
accelerating cosmic rays then our comparison of the electron
temperature with the shock velocity tells us nothing about \tetp . In
fact, non-linear models of shock acceleration \citep{Ellison00}, at
Mach numbers appropriate for 1E~0102$-$72, exhibit electron
temperatures near 1~keV and fully equilibrated electron and proton
thermal populations.

\section{Conclusions \label{s_finale}}

\begin{figure}
\centerline{\includegraphics[scale=0.9, keepaspectratio]
{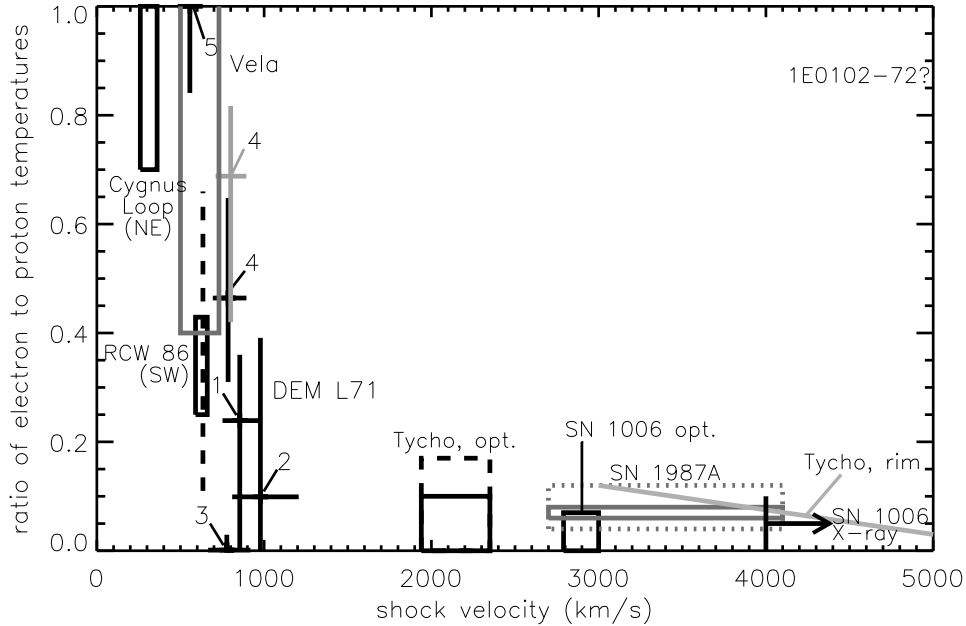}}
\caption{Compilation of current electron-ion equilibration
measurements at SNR shocks. The solid boxes for the Cygnus Loop, 
RCW 86 and Tycho, are from \citet{Ghav01} modeling the broad-to-narrow 
ratios of H$\alpha$
and H$\beta$. The dashed lines are my estimate using only the H$\beta$ 
measurement, however this estimate does not include the potentially
significant effect of precursor contamination to the narrow component.
The grey box marked ``Vela'' shows the estimate by \citet{Bocchino99}
of the current \tetp , which did not take
into account the amount of coulomb equilibration that has already 
occurred.
Crosses show DEM L71 results from \citet{Rak03} using the broad
H$\alpha$ component to fix the proton temperature in an X-ray fit of
\tetp . The grey cross indicates the updated result
from deeper observations which allow us to simultaneously constrain 
the elemental abundances rather than fixing them to some 
average abundance set. 
The two SN 1006 results indicate the most constraining of the
optical or UV line ratio measurements \citep{Ghav03}, and the X-ray
grating measurement of the oxygen line width and electron
temperature \citep{Vink03}. The solid box for SN 1987A shows the
reported results from \citet{Michael02} using a fixed shock velocity
to constrain \tetp from the X-ray CCD spectrum, the dashed box gives an
estimate which includes the error in their measurement of the shock
velocity. The \citet{Hwang02} measurement plotted here, converted from
their electron-to-mean temperature ratio for Tycho's outer rim,
compares the electron temperature measurement with both the radio and
X-ray expansion velocities. 
For 1E~0102$-$72, the low electron temperature could not be explained
by a minimal degree of electron-ion equilibration at the shock
front but rather required that 
a significant portion of the shock energy be going into the
acceleration of cosmic rays. If the injection efficiency into the
acceleration process is high, models predict that the remaining
thermal populations may be equilibrated.}
\label{finale}
\end{figure}

The preceding results are compiled in figure \ref{finale}.
Note the new constraint for region 4 in DEM~L71, obtained using a
\chandra\ observation three times as long as the original. 
This region still requires an
intermediate degree of equilibration even without assuming a
particular abundance set.
Recall also that corroborating ion-ion equilibration measurements exist
for SN 1006\citep[mass-proportional heating]{Ray95} and the Cygnus
Loop \citep[$T_{oxygen} < 2.5 \times T_{\rm p}$]{Ray03}. 
Also included in this figure is the report 
by \citet{Bocchino99} of \tetp\ between 0.4 and 1 
for an approximately 600 km s$^{-1}$ shock in the
Vela SNR using ROSAT PSPC spectra. This was estimated by 
comparing the electron temperature of
the hotter of two collisional ionization equilibrium components 
with a proton temperature derived from their
best estimates for the Sedov parameters for Vela.  
However a measurement of the ionization timescale is needed to
calibrate the extent to which coulomb collisions have already
equilibrated the electron and proton temperatures as was done for
DEM~L71 and 1E~0102$-$72.
  
An inverse correlation between shock velocity and efficiency of electron
heating has been suggested based on different subsets of these
results. However, even if we accept that the high equilibration result
for the Cygnus Loop is not an artifact of neglecting the pre-cursor
emission, we are still left with one secure identification of
an intermediate equilibration in the interesting range from 400 to
2000 km s$^{-1}$ where the transition from near full equilibration to
low equilibration should occur.
Furthermore, once cosmic ray acceleration
is considered, the shock velocities that were inferred from thermal
line-widths may be grossly underestimated. 
Additionally, the fast shocks in Tycho,
while not requiring a large cosmic ray population, do not exclude it
either. If the outer blast-wave of Tycho is efficiently accelerating
particles to cosmic ray energies, synchrotron radiation from high 
energy electrons may explain the featureless rim spectra while the
thermal populations could be nearly equilibrated similarly to 
1E~0102$-$72. 

Progress in this developing field depends on accurate modeling of the
emission from pre-shock and post-shock gas, corroborating evidence
from multiple wave-bands, and useful assessments of the cosmic ray
production and its effect on the shock. 
The spectral resolution of future X-ray missions such as ASTRO-E2,
ConX and Xeus will allow thermal Doppler broadening to be measured 
routinely in the X-rays as they now are in the optical and
ultraviolet. 
Observations of 500-2000 km
s$^{-1}$ shocks, where the transition from full to minimal
equilibration should occur are especially important for diagnosing the
relationship between shock velocity and the relative importance of
different collisionless heating processes.

I would like to thank Arend Sluis, Parviz Ghavamian and Kelly Korreck
for fruitful discussions regarding this review and shock physics in
general. This review also benefited from the comments of two
conscientious referees. This work was supported by NASA grant NAG5-9281. 



\begin{thebibliography}{}


\bibitem[Bocchino et al.(1999)]{Bocchino99}
Bocchino, F., Maggio, A., and Sciortino, S. ``ROSAT PSPC observation
of the NE region of the Vela supernova remnant III. The two-component
nature of the X-ray emission and its implications on the ISM''
Astron. Astrophys. 342, 839-853, 1999.

\bibitem[Chevalier et al.(1980)Chevalier, Raymond, and
Kirshner]{CRK80}
Chevalier, Roger. A., Kirshner, Robert. P., and Raymond, John. C.
``The optical emission from a fast shock wave with application to
supernova remnants''. ApJ 235, 186-195, 1980.

\bibitem[Decourcelle et al.(2000)Decourchelle, Ellison, and
  Ballet]{DEC00}
Decourchelle, Anne, Ellison, D. C., and Ballet, Jean
``Thermal X-Ray Emission and Cosmic-Ray Production in Young Supernova
  Remnants'' ApJ 543, L57-L60, 2000.

\bibitem[Ellison(2000)]{Ellison00}
Ellison, D. C. ``The Cosmic Ray-X-ray Connection: Effects of Nonlinear
Shock Acceleration on Photon Production in SNRs'' in AIP
Conf. Proc. 528, Acceleration and Transport of Energetic Particles
Observed in the Heliosphere.  
Edited by Richard A. Mewaldt, J. R. Jokipii, Martin A. Lee, 
Eberhard M\"obius, and Thomas H. Zurbuchen. Melville, (New York: AIP), 
pp. 386-93, 2000. 

\bibitem[Ghavamian et al.(2001)]{Ghav01}
Ghavamian, Parviz, Raymond, John C., Smith, R. Chris, and Hartigan,
Patrick ``Balmer-dominated spectra of nonradiative shocks in the
Cygnus Loop, RCW 86, and Tycho supernova remnants.'' ApJ 547,
995-1009, 2001.

\bibitem[Ghavamian et al.(2002)]{Ghav02}
Ghavamian, Parviz, Winkler, P. Frank, Raymond, John C., and Long,
Knox~S. ``The optical spectrum of the SN 1006 supernova remnant
revisited.'' ApJ, 572, 888-896, 2002.

\bibitem[Ghavamian et al.(2003)]{Ghav03}
Ghavamian, Parviz, Rakowski, Cara E., and Hughes, John P. ``The physics
of supernova remnant blast waves. I. Kinematics of DEM L71 in the
Large Magellanic Cloud.'' ApJ 590, 833-845, 2003. 

\bibitem[Hughes(2000)]{Hughes00a}
Hughes, John P. ``The Expansion of the X-ray Remnant of Tycho's 
Supernova (SN 1572)'' ApJ, 545, L53-L56, 2000.

\bibitem[Hughes et al.(2000)Hughes, Rakowski, and Decourchelle]{Hughes00b}
Hughes, John P., Rakowski, Cara E., Decourchelle, Anne ``Electron
heating and cosmic rays at a supernova shock from {\it Chandra} X-ray
observations of 1E 0102.2$-$7219.'' ApJ 543, L61-L65, 2000.

\bibitem[Hwang et al.(2002)]{Hwang02}
Hwang, Una,  Decourchelle, Anne,  Holt, Stephen, S.,  and Petre, Robert  
``Thermal and nonthermal X-ray emission from the forward shock in Tycho's
supernova remnant.'' ApJ 581, 1101-1115, 2002.

\bibitem[Laming et al.(1996)]{Laming96}
Laming, J. Martin, Raymond, John C., McLaughlin, Brendan M. and Blair,
William P.  ``Electron-ion equilibration in nonradiative shocks associated with SN 1006.'' ApJ 472, 267-274, 1996. 

\bibitem[Lembege et al.(2004)]{Lem04}
Lembege, B., Giacalone, J., Scholer, M., et al. ``Selected problems in
collisionless-shock physics'' Space Science Reviews 110, 161-226, 2004.


\bibitem[McKee and Hollenbach(1980)]{McKee80}
McKee, Christopher F., and Hollenbach, David J. ``Interstellar Shock
Waves'' Ann. Rev. Astron. Astrophys. 18, 219-262, 1980.

\bibitem[Michael et al.(2002)]{Michael02}
Michael, Eli,  Zhekov, Svetozar,  McCray Richard, et al. ``The X-ray
spectrum of supernova remnant 1987A.'' ApJ 574, 166-178, 2002.

\bibitem[Rakowski et al.(2003)Rakowski, Ghavamian and Hughes]{Rak03}
Rakowski, Cara E., Ghavamian, Parviz, and Hughes, John P. ``The physics
of supernova remnant blast waves. II. Electron-ion equilibration in
DEM L71 in the Large Magellanic Cloud.'' ApJ 590, 846-857, 2003.

\bibitem[Raymond et al.(1995)Raymond, Blair, and Long]{Ray95}
Raymond, John C., Blair, William P., and Long, Knox S. ``Detection of
ultraviolet emission lines in SN 1006 with the Hopkins Ultraviolet
Telescope.'' ApJ 454, L31-L34, 1995.

\bibitem[Raymond et al.(2003)]{Ray03} 
Raymond, John C.,  Ghavamian, Parviz,  Sanskrit, Ravi, et
al. ``Far-ultraviolet spectra of a nonradiative shock wave in the
Cygnus Loop.'' ApJ 584, 770-781, 2003. 

\bibitem[Reynoso et al.(1997)]{Rey97}
Reynoso, E. M., Moffett, D. A., Goss, W. M., et al. ``A VLA Study of
the Expansion of Tycho's Supernova Remnant'' ApJ 491, 816-828, 1997.

\bibitem[Schwartz et al.(1988)]{Schwartz88}
Schwartz, Steven. J., Thomsen Michelle. F., Bame, S. J., and Stansberry,
John ``Electron heating and the potential jump across fast mode
shocks'' J. Geophys. Rev. 93, 12923-12931, 1988.

\bibitem[Vink et al.(2003)]{Vink03}
Vink, Jacco, Laming, J. Martin, Go, Ming Feng, et al. ``The slow
temperature equilibration behind the shock front of SN 1006.'' ApJ
587, L31-L34, 2003.



\end{thebibliography}
\end{document}